\documentclass[12pt]{article}

\usepackage[left=1in, right=1in, top=1in, bottom=1in]{geometry}
\usepackage{amsmath, amssymb, amsthm}
\usepackage{graphicx}
\usepackage{natbib}
\usepackage{bm}
\theoremstyle{definition}

\newcommand{\h}{\mathcal{H}}

\DeclareMathOperator{\BIC}{BIC}
\DeclareMathOperator{\BF}{BF}
\DeclareMathOperator{\PBF}{PBF}

\def\spacingset#1{\renewcommand{\baselinestretch}%
{#1}\small\normalsize} \spacingset{1}

\begin{document}
\title{\bf Closed-form approximations of the two-sample Pearson Bayes factor}

\author{Thomas J. Faulkenberry\thanks{Corresponding author: faulkenberry@tarleton.edu}\hspace{.2cm}\\
    \small{Department of Psychological Sciences, Tarleton State University,
    Stephenville, TX, USA}}

\maketitle

\bigskip

\begin{abstract}
  In this paper, I present three closed-form approximations of the two-sample Pearson Bayes factor, a recently developed index of evidential value for data in two-group designs. The techniques rely on some classical asymptotic results about gamma functions. These approximations permit simple closed-form calculation of the Pearson Bayes factor in cases where only minimal summary statistics are available (i.e., the $t$-score and degrees of freedom). Moreover, these approximations vastly outperform the classic BIC method for approximating Bayes factors from experimental designs.
\end{abstract}

\noindent%

\noindent%
{\it Word count:} 3,126
\vfill

\newpage
\spacingset{1.3} % DON'T change the spacing!

 A common scenario in applied statistical inference involves comparing the means of two independent samples (e.g., a treatment group and a control group). This can be done using hypothesis testing (or more broadly, model comparison), where two hypotheses about potential differences between the two underlying population means $\mu_1, \mu_2$ are compared after observing data. Classically, hypothesis tests work by first assuming a {\it null hypothesis} $\h_0$, and then calculating a test statistic that can be used to index the probability of obtaining some sample of observed data under the null hypothesis. If this probability is small, the common logic is to reject the null hypothesis $\h_0$ in favor of some {\it alternative hypothesis} $\h_1$. The problem tackled in the present paper is twofold. First, we consider an alternative method for this type of inference based on {\it Bayes factors} \citep{kass1995,faulkenberryBayes} and provide a novel, but simple method for their computation. Second, we compare this method with a classic approximation based on the Bayesian information criterion (BIC) and show that our new method outperforms the BIC method by a significant margin.

Classically, inference about potential differences between two group means is done with the \(t\)-test. To begin, we will review some background on the \(t\)-test. Accordingly, let us consider an experiment with two groups each containing $N$ independent measurements. Let $y_{ij}$ denote the $i^{th}$ measurement ($i=1,\dots,N$) in the $j^{th}$ group ($j=1,2$). Further, we assume that the $y_{ij}$ are drawn from independent and normally distributed populations with mean $\mu_j$ and variance $\sigma^2$. Then we can test the hypotheses%
\[
  \h_0:\mu_1=\mu_2\text{  versus  }\h_1:\mu_1\neq \mu_2
\]%
by computing a test statistic%
\[
  t=\frac{\overline{y}_1-\overline{y}_2}{\hat{\sigma}_p/\sqrt{N_{\delta}}} \;.
\]%
Here, $\overline{x}_j$ represents the sample mean of the measurements in group $j$ and $\hat{\sigma}_p$ is the pooled estimate of $\sigma$, defined by the relationship%
\[
  \hat{\sigma}^2_p = \frac{\hat{\sigma}_1^2(N_1-1)+\hat{\sigma}_2^2(N_2-1)}{N_1+N_2-2},
\]%
where each $\hat{\sigma}_j$ is the sample standard deviation of the measurements in group $j$. Finally, $N_{\delta} = (1/N_1+1/N_2)^{-1}$, a quantity often referred to as the {\it effective sample size} for the experiment. 

Under the null hypothesis $\mathcal{H}_0$, the distribution of these $t$ scores is well known as {\sl Student's $t$ distribution}, a random variable (denoted $T_{\nu}$) with density function
\[
  f_{\nu}(x) = \frac{\Gamma\left(\frac{\nu + 1}{2}\right)}{\sqrt{\nu \pi} \Gamma\left(\frac{\nu}{2}\right)} \left(1+\frac{x^2}{\nu}\right)^{-\frac{\nu+1}{2}},
\]
where $\nu=N_1+N_2-2$ represents the {\it degrees of freedom} of the test. The cumulative distribution function $F_{\nu}(x)=\int_{-\infty}^x f_{\nu}(u)du$ is used to index the probability of observing data at least as extreme as that which we observed under the null hypothesis $\h_0$. Specifically, for an observed $t$-statistic $t_{\text{obs}}$, we compute $P(|T_\nu|>t_{\text{obs}}) = 2(1-F_{\nu}(t_{\text{obs}}))$, a quantity commonly known as a $p$-value.  If this \(p\)-value is small (say, less than 5\%), the typical decision is to {\it reject} $\mathcal{H}_0$ in favor of $\h_1$ and conclude that $\mu_1\neq \mu_2$, thus implying that the two populations from which we sampled indeed have different underlying means.

Despite the popularity of this approach to testing for differences in sample means, there have been many recent criticisms against their use, and more generally, against null hypothesis significance testing \citep{wagenmakers2007}. Additionally, the American Statistical Association has recently recommended against the use of $p$-values and significance testing for scientific inference \citep{asa}. One alternative that has been recommended is a {\it Bayesian} approach, which works by considering the {\it relative predictive adequacy} of $\h_0$ and $\h_1$ against some observed data $\bm{y}$. This relative predictive adequacy is indexed by computing the {\it Bayes factor} \citep{kass1995}, which indexes the extent to which the observed data $\bm{y}$ is more likely under one hypothesis -- say $\h_1$ -- compared to the other hypothesis -- say $\h_0$. That is,%
\[
  \BF_{10} = \frac{p(\bm{y} \mid \h_1)}{p(\bm{y} \mid \h_0)}
\]%
where $p(\bm{y} \mid \h_i)$ is the {\it marginal likelihood} of $\bm{y}$ under $\h_i$, defined as%
\[
  p(\bm{y} \mid \h_i) = \int p(\bm{y} \mid \theta_i,\h_i)\cdot \pi(\theta_i,\h_i)d\theta_i.
\]%

In general, computing Bayes factors can be quite difficult, particularly because computing the marginal likelihoods involves nontrivial integration. One popular approach adopted in recent years is the {\it BIC approximation} \citep{kass1995, wagenmakers2007, masson2011}, which works by constructing a second-order Taylor approximation of the log marginal likelihood about the posterior mode for $\theta_i$. \citet{kass1995} show that this results in the approximation%
\[
  p(\bm{y}\mid \h_i) \approx \exp\Bigl(-\frac{1}{2}\BIC(\h_i)\Bigr) \;.
\]%
This approximation uses the Schwarz Bayesian Information Criterion (BIC) \citep{schwarz1978}, which is computed as $\BIC(\h_i)=-2\ln L_i + k_i\ln n$, where $k_i$ is the number of parameters in $\mathcal{H}_i$, $n$ is the number of data observations, and $L_i$ is the value of the likelihood function for model $\mathcal{H}_i$ with the maximum likelihood estimate taken as its argument. As typically used (e.g., our two-sample hypothesis testing problem), $n$ is taken as the number of distinct observations across all experimental groups (i.e., $n=2N$) and $k_1 = k_0+1$. This last equality represents the fact that $\h_1$ has an additional parameter indexing the experimental group that $\h_0$ does not have. From here, it is easy to derive an approximation for the Bayes factor:
\begin{align*}
  \BF_{01} = \frac{p(\bm{y}\mid \h_0)}{p(\bm{y}\mid \h_1)} & \approx \frac{\exp\Bigl(-\frac{1}{2}\BIC(\h_0)\Bigr)}{\exp\Bigl(-\frac{1}{2}\BIC(\h_1)\Bigr)}\\
  &= \exp \Biggl(\frac{\BIC(\h_1) - \BIC(\h_0)}{2}\Biggr).
\end{align*}

Compared to computing marginal likelihoods by integrating the prior-weighted likelihoods, the BIC approximation requires only that we know the BIC values for models $\h_0$ and $\h_1$. Generally, this requires that we have ``raw'' data available. Recent work by \citet{faulkenberry2018} improved the BIC method by making the Bayes factor computation accessible with only summary statistics:%
\[
  \BF_{01} \approx \sqrt{n\Biggl(1+\frac{t^2}{\nu}\Biggr)^{-n}}.
\]

Despite the simplicity of computation offered by the BIC method, the fact remains that it is based on a large sample approximation, leaving it with limited utility in a small sample context. It is also limited for use in a teaching context, where sample sizes for hand-worked examples tend to be small. Recent work by \citet{wangSun} has provided another approach for calculating Bayes factors that can prove to be fruitful. In their work, Wang and Sun place a random effects linear model on the data%
\[
  y_{ij}=\mu + a_j + \varepsilon_{ij},
\]%
assuming that $a_j\sim \mathcal{N}(0,\sigma^2_a)$ and $\varepsilon_{ij} \sim \mathcal{N}(0,\sigma^2)$ (for $i=1,\dots,N$, $j=1,2$). Cast in this form, the goal of the hypothesis test is to test whether the random effects term $a_j$ is identically 0. We define the null hypothesis by restricting the variability of the random effects term, giving%
\[
  \h_0:\sigma^2_a=0 \text{ versus }\h_1:\sigma_a^2\neq 0.
  \]

\begin{figure}[h]
  \centering
  \includegraphics[width=0.8\textwidth]{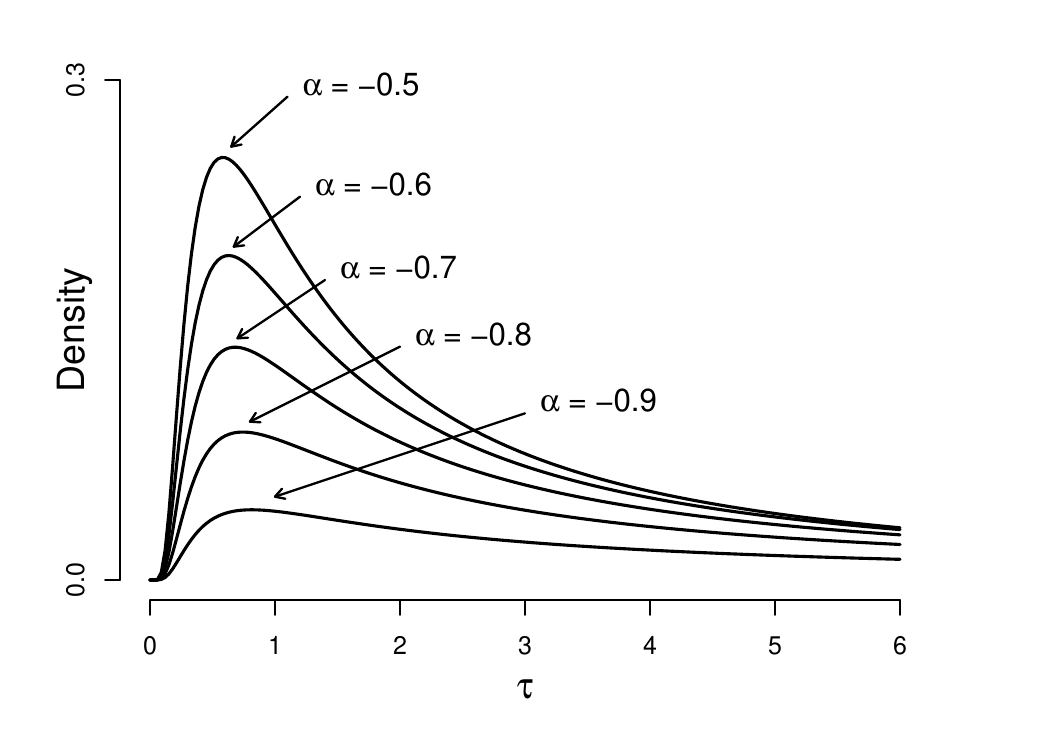}
  \caption{A Pearson Type VI prior for $\tau$, plotted as a function of shape parameter $\alpha$.}
  \label{fig:prior}
\end{figure}

Wang and Sun applied a method of \citet{gds}, who considered a proper prior on the ratio of variance components $\tau = \sigma^2_a/\sigma^2$ under $\h_1$. Under this prior specification, Garcia-Donato and Sun showed that the Bayes factor could be computed as%
\[
  \BF_{10} = \int_0^{\infty}(1+\tau r)^{\frac{1-p}{2}}\Biggl(1-\frac{\tau r}{1+\tau r}\cdot \frac{SSA}{SST}\Biggr)^{\frac{1-pr}{2}}\cdot \pi(\tau)d\tau
\]%
where $p$ = number of experimental groups, $r$ = number of replicates per group, $SST$ represents the total variability of the data (i.e., ``total sum of squares'', $SST = \sum_i \sum_j (y_{ij} - \overline{y}_{\cdot \cdot})^2)$ and $SSA$ represents the variability due to differences between the two experimental groups (i.e., $SSA=n\sum_j(\overline{y}_{\cdot j} - \overline{y}_{\cdot \cdot})^2)$). To use this equation, the user must place a prior distribution on the variance components $\tau$. Wang and Sun used a {\it Pearson Type VI distribution} to serve as the prior for $\tau$. The Pearson Type VI distribution has three parameters: two shape parameters $\alpha>-1$ and $\beta>-1$ and a scale parameter $\kappa>0$. The density function for the Pearson Type VI prior (defined for $\tau>0$) is given by%
\[
  \pi^{PT}(\tau) = \frac{\kappa(\kappa \tau)^{\beta}(1+\kappa \tau)^{-\alpha-\beta-2}}{\mathcal{B}(\alpha+1, \beta+1)}I_{(0,\infty)}(\tau)
\]%
where $\mathcal{B}(x,y) = \int_0^1t^{x-1}(1-t)^{y-1}dt$ is the standard Beta function. \citet{wang2016} further restricted $\pi^{PT}$ to one parameter $\alpha \in [-1,-\frac{1}{2}]$ by taking $\kappa=r$ and $\beta = \frac{n-p}{2}-\alpha-2$. A plot of this prior can be seen in Figure \ref{fig:prior}; in this figure, we assume $p=2$ groups and $r=20$ replicates per condition, giving a total of $n=40$ observations, thus setting $\kappa=20$ and $\beta=\frac{40-2}{2}-\alpha-2$, where $\alpha$ takes specific values between -1 and -1/2. Further, Figure \ref{fig:prior} shows the effect of varying $\alpha$ on the prior distribution for $\tau$. As we can see, as $\alpha$ decreases from -1/2 to -1, $\tau$ becomes more dispersed and less peaked around the mode. This places more prior mass on larger treatment effects than we would see for values of $\alpha$ closer to $-\frac{1}{2}$.

With the above prior specification, Wang and colleagues \citep{wangSun,wang2016}  proved that the Bayes factor of Garcia-Donato and Sun simplifies to an analytic expression without integral representation. The resulting {\it Pearson Bayes factor} \citep{faulkenberryPearson} allows one to calculate the Bayes factor for $\h_1$ over $\h_0$ without the need for integration:%
\[
  \PBF_{10} = \frac{\Gamma\Bigl(\frac{\nu}{2}\Bigr) \cdot \Gamma\Bigl(\alpha + \frac{3}{2}\Bigr)}{\Gamma\Bigl(\frac{\nu+1}{2}\Bigr)\cdot \Gamma(\alpha+1)}\Biggl(1+\frac{t^2}{\nu}\Biggr)^{(\nu-2\alpha-2)/2}.
\]%
The Pearson Bayes factor is computable using only the summary statistics from the \(t\)-test (i.e., the $t$-score and the degrees of freedom $\nu$). It also includes a parameter $\alpha$ which allows the analyst to tune the scale of the prior distribution on effect sizes. \citet{wang2016} recommend a default setting of $\alpha=-1/2$, as the asymptotic tail behavior of the prior becomes a Cauchy distribution as $\tau$ increases, mirroring the prior structure recommended by \citet{jeffreys1961}. Following this recommendation, the Pearson Bayes factor simplifies to:%
\begin{equation}\label{eq:pbf}
  \PBF_{10} = \frac{\Gamma\Bigl(\frac{\nu}{2}\Bigr) \cdot \Gamma(1)}{\Gamma\Bigl(\frac{\nu+1}{2}\Bigr)\cdot \Gamma\Bigl(\frac{1}{2}\Bigr)}\Biggl(1+\frac{t^2}{\nu}\Biggr)^{(\nu-1)/2} = \frac{\Gamma\Bigl(\frac{\nu}{2}\Bigr)}{\Gamma\Bigl(\frac{\nu}{2}+\frac{1}{2}\Bigr)}\sqrt{\frac{1}{\pi}\Biggl(1+\frac{t^2}{\nu}\Biggr)^{\nu-1}}.
\end{equation}

While Equation \ref{eq:pbf} seems relatively simple to compute, its reliance on the Gamma function $\Gamma(x)$ is problematic for two reasons. The first reason is that the Gamma function involves integration:

\[
  \Gamma(x) = \int_0^{\infty} t^{x-1}e^{-t}dt \; .
\]%
In situations where only a scientific calculator may be available (i.e., a common situation in teaching), users will have no easy way to compute the Gamma functions, thus leaving those without sufficient mathematical background to be potentially deterred from using the formula. The second reason is that in statistical computing languages, the output of the Gamma function quickly exceeds the maximum value of double-precision machine number representation (about $1.8 \times 10^{308}$). For example, in R, the largest computable integer argument for the Gamma function is 171; anything larger than 172 returns \verb|Inf|. Thus, directly computing the Pearson Bayes factor in R via Equation \ref{eq:pbf} will fail whenever the degrees of freedom $\nu$ is greater than 343 (i.e., a combined sample size greater than 345).

For these reasons, it is desirable to find {\it closed-form} approximations of the Pearson Bayes factor that will mitigate these problems, rendering the formula computable for large samples, as well as potentially increasing its accessibility to a broader audience of users. In this paper, I will present three such approximations, demonstrate their use, and compare them with each other.

To this end, the main work of this paper concerns the following. Let us first rewrite Equation \ref{eq:pbf} as%
\begin{equation}\label{eq:pbf2}
  \PBF_{10} = C_{\nu}\cdot \sqrt{\frac{1}{\pi}\Biggl(1+\frac{t^2}{\nu}\Biggr)^{\nu-1}} \text{ where } C_{\nu} = \frac{\Gamma\Bigl(\frac{\nu}{2}\Bigr)}{\Gamma\Bigl(\frac{\nu}{2}+\frac{1}{2}\Bigr)}.
\end{equation}%
Our goal is to find closed-form approximations of the constant $C_{\nu}$ that can be computed using only elementary functions (i.e., with a simple scientific calculator), and then compare these approximations with the BIC approximation.

\section{Approximating quotients of Gamma functions}
Against the background of the previous section, we are ready to tackle the problem at hand. As we just demonstrated, computing Bayes factors directly from observed $t$-scores requires being able to compute the quotient%
\[
  C_{\nu} = \frac{\Gamma\Bigl(\frac{\nu}{2}\Bigr)}{\Gamma\Bigl(\frac{\nu}{2}+\frac{1}{2}\Bigr)}.
\]%
Direct computation of these Gamma functions requires calculus (or more practically, numerical routines in a scientific programming language). Thus, the goal in this paper is to find closed-form approximations of this quotient that can be carried out using only basic algebraic operations. To this end, I have developed three such approximations -- one that follows directly from a classical asymptotic formula of \citet{wendel}, one that derives directly from the classical Stirling formula \citep{jameson}, and finally, one that follows from an ``improved'' approximation of \citet{frame}. 
    
\subsection{Wendel's asymptotic formula}

In his brief paper, \citet{wendel} showed that for all real numbers $a$ and $x$,%
\[
  \frac{\Gamma(x+a)}{x^a\Gamma(x)} \approx 1 \text{ as } x\rightarrow \infty \; .
\]%
or equivalently,%
\[
  \frac{\Gamma(x)}{\Gamma(x+a)} \approx \frac{1}{x^a} \text{ as } x\rightarrow \infty ;.
\]%
Letting $x=\frac{\nu}{2}$ and setting $a=1/2$, we cast Wendel's formula into a form that proves useful for our current problem of approximating $C_{\nu}$;
\[
  C_{\nu} = \frac{\Gamma\Bigl(\frac{\nu}{2}\Bigr)}{\Gamma\Bigl(\frac{\nu}{2} + \frac{1}{2}\Bigr)} \approx \frac{1}{(\nu/2)^{1/2}} = \sqrt{\frac{2}{\nu}} \; .
\]%
Thus, we can combine this with Equation \ref{eq:pbf2} to immediately derive the following approximation for the two-sample Pearson Bayes factor:%
\[
  \PBF_{10} \approx \sqrt{\frac{2}{\nu}}\cdot \sqrt{\frac{1}{\pi}\Biggl(1+\frac{t^2}{\nu}\Biggr)^{\nu-1}} = \sqrt{\frac{2}{\pi\nu}\Biggl(1+\frac{t^2}{\nu}\Biggr)^{\nu-1}} \; .
\]

\subsection{Stirling's formula}

Another approach to approximating $C_{\nu}$ comes from applying Stirling's formula \citep{jameson}. Historically, Stirling's formula arose as a way to approximate the factorial function for the positive integers; i.e.,%
\[
  n! \approx \sqrt{2\pi}n^{n+\frac{1}{2}}e^{-n} \; .
\]%
As the Gamma function $\Gamma(x)$ can be seen as a continuous extention of the factorial function, it is natural to extend Stirling's formula to hold for any real number $x$, not just positive integers. In fact, this extension is reasonably easy to predict (just note that for positive integer $n$, $\Gamma(n) = (n-1)!$):%
\begin{equation}\label{eq:stirling}
  \Gamma(x) \approx \sqrt{2\pi}x^{x-\frac{1}{2}}e^{-x} \; .
\end{equation}%
Thus, it is easy to use Equation \ref{eq:stirling} to compute a closed form approximation for $C_{\nu}$. To this end, it is straightforward to show%
\begin{align*}
  C_{\nu} &= \frac{\Gamma\Bigl(\frac{\nu}{2}\Bigr)}{\Gamma\Bigl(\frac{\nu}{2} + \frac{1}{2}\Bigr)}\\
          & \approx \frac{\sqrt{2\pi}\cdot \Bigl(\frac{\nu}{2}\Bigr)^{\frac{\nu}{2}-\frac{1}{2}}\cdot e^{-\frac{\nu}{2}}}{\sqrt{2\pi}\cdot \Bigl(\frac{\nu}{2}+\frac{1}{2}\Bigr)^{\frac{\nu}{2}+\frac{1}{2}-\frac{1}{2}}\cdot e^{-\frac{\nu}{2}-\frac{1}{2}}}\\
  &= \sqrt{\frac{2e}{\nu}}\Biggl(\frac{\nu}{\nu+1}\Biggr)^{\nu/2} \; .
\end{align*}%
Combining this with Equation \ref{eq:pbf2} gives another approximation for the two-sample Pearson Bayes factor:%
\[
  \PBF_{10} \approx \sqrt{\frac{2e}{\pi(\nu+1)}}\Biggl(\frac{\nu+t^2}{\nu+1}\Biggr)^{(\nu-1)/2} \; .
\]

\subsection{Frame's quotient formula}

The final approach I will explore in this paper is derived from a method of \citet{frame}, who proposed the following approximation to the quotient of two nearby values of the Gamma function:%
\begin{equation}\label{eq:frame}
  \frac{\Gamma\Bigl(n+\frac{1+u}{2}\Bigr)}{\Gamma\Bigl(n+\frac{1-u}{2}\Bigr)} \approx \Biggl(n^2 + \frac{1-u^2}{12}\Biggr)^{\frac{u}{2}}\; .
\end{equation}%
To apply the Frame approximation, we must first transform the left hand side of Equation \ref{eq:frame} into a form more appropriate for computing $C_{\nu}$. The critical step is to set%
\[
  u=-\frac{1}{2} \text{ and } n = \frac{2\nu-1}{4} \; .
\]%
The reader can easily verify that this indeed works:%
\[
  \frac{\Gamma\Bigl(n+\frac{1+u}{2}\Bigr)}{\Gamma\Bigl(n+\frac{1-u}{2}\Bigr)}  =   \frac{\Gamma\Bigl(\frac{2\nu-1}{4}+\frac{1-\frac{1}{2}}{2}\Bigr)}{\Gamma\Bigl(\frac{2\nu-1}{4}+\frac{1+\frac{1}{2}}{2}\Bigr)} = \frac{\Gamma\Bigl(\frac{\nu}{2}\Bigr)}{\Gamma\Bigl(\frac{\nu}{2}+\frac{1}{2}\Bigr)}= C_{\nu} \; .
\]%
Thus, using Frame's approximation gives%
\begin{align*}
  C_{\nu} &\approx \Biggl(n^2 + \frac{1-u^2}{12}\Biggr)^{\frac{u}{2}}\\
          & = \Biggl(\Bigl(\frac{2\nu-1}{4}\Bigr)^2 + \frac{1-(-1/2)^2}{12}\Biggr)^{-\frac{1/2}{2}}\\
          &= \Biggr(\frac{4\nu^2-4\nu+1}{16} + \frac{1}{16}\Biggr)^{-\frac{1}{4}}\\
          &= \Biggr(\frac{8}{2\nu^2-2\nu+1}\Biggr)^{\frac{1}{4}}\; .
\end{align*}

Combining this with Equation \ref{eq:pbf2} gives a third closed-form approximation for the two-sample Pearson Bayes factor:%
\[
  \PBF_{10} \approx \Biggr(\frac{8}{2\nu^2-2\nu+1}\Biggr)^{\frac{1}{4}} \cdot \sqrt{\frac{1}{\pi}\Biggl(1+\frac{t^2}{\nu}\Biggr)^{\nu-1}} \; .
\]  

\section{Example computations}

For illustration, let us now apply these three approximation methods to a concrete example. Consider the following summary data from \citet{borota2014}, who observed that with a sample of $n=73$ participants, those who received 200 mg of caffeine performed significantly better on a test of object memory compared to a control group of participants who received a placebo, $t(71)=2.0$, $p=0.049$. Borota and colleagues claimed this result as evidence that caffeine enhances memory consolidation.

First, we apply the Wendel approximation. Using the summary data from Borota et al. gives%
\begin{align*}
  \PBF_{10} & \approx \sqrt{\frac{2}{\nu}}\cdot \sqrt{\frac{1}{\pi}\Biggl(1+\frac{t^2}{\nu}\Biggr)^{\nu-1}}\\
            & \sqrt{\frac{2}{71}} \cdot \sqrt{\frac{1}{\pi}\Biggl(1+\frac{2.0^2}{71}\Biggr)^{71-1}}\\
            &= 0.1678 \cdot 3.8417\\
            &= 0.6446 \; .
\end{align*}%
This value of the Bayes factor implies that Borota et al.'s data are $\PBF_{01} = 1/\PBF_{10} = 1/0.6446 = 1.551$ times more likely under the null hypothesis $\mathcal{H}_0$ than under the alternative hypothesis $\mathcal{H}_1$, thus giving positive evidence for caffeine having a null effect on memory consolidation. 

Note that this calculation can be done using only a simple scientific calculator. How does it compare to the analytic (i.e., non-approximated) Pearson Bayes factor? If we use Equation \ref{eq:pbf2} and calculate $C_{\nu}$ analytically, we get%
\begin{align*}
  \PBF_{10} & = C_{\nu}\cdot \sqrt{\frac{1}{\pi}\Biggl(1+\frac{t^2}{\nu}\Biggr)^{\nu-1}}\\
            &= \frac{\Gamma\Bigl(\frac{\nu}{2}\Bigr)}{\Gamma\Bigl(\frac{\nu}{2}+\frac{1}{2}\Bigr)}\cdot \sqrt{\frac{1}{\pi}\Biggl(1+\frac{t^2}{\nu}\Biggr)^{\nu-1}}\\
            &= \frac{\Gamma\Bigl(\frac{71}{2}\Bigr)}{\Gamma\Bigl(\frac{71}{2}+\frac{1}{2}\Bigr)}\cdot \sqrt{\frac{1}{\pi}\Biggl(1+\frac{2.0^2}{71}\Biggr)^{71-1}}\\
            &= 0.1684 \cdot 3.8417\\
  &= 0.6469 \; .
\end{align*}%
The approximation error we incur by using the Wendel asymptotic formula for approximating $C_{\nu}$ is small, resulting in an underestimate of $0.6446-0.6469 = -0.0023$, a relative error magnitude of 0.36\%. For comparison, consider the error that results from using the BIC method \citep{kass1995,wagenmakers2007,masson2011}, a popular method for approximating Bayes factors direclty from summary statistics. \citet{faulkenberry2018} showed that the BIC Bayes factor can be computed directly as follows:%
\begin{align*}
  \text{BF}_{01} & \approx \sqrt{n\cdot \Bigl(1 + \frac{t^2}{\nu}\Bigr)^{-n}}\\
                 &= \sqrt{73 \cdot \Bigl(1 + \frac{2.0^2}{71}\Bigr)^{-73}}\\
  &= 1.1557 \; .
\end{align*}%
Keeping in mind that the BIC Bayes factor expresses evidence for $\mathcal{H}_0$, we reciprocate to compute $\text{BF}_{10} = 1/\text{BF}_{01} = 0.8653$. Compared to the analytic Pearson Bayes factor, this is a overestimate of $0.8653-0.6469 = 0.2184$, relative error magnitude of 33.7\%. Our new method based on Wendel's asymptotic approximation of the Gamma function improves on this error by two orders of magnitude.

Next, we will now apply the approximation based on Stirling's formula to the \citet{borota2014} summary statistics. This yields%
\begin{align*}
  \PBF_{10} & \approx \sqrt{\frac{2e}{\pi(\nu+1)}}\Bigl(\frac{\nu+t^2}{\nu+1}\Bigr)^{(\nu-1)/2}\\
            &= \sqrt{\frac{2e}{\pi(71+1)}}\Bigl(\frac{71+2.0^2}{71+1}\Bigr)^{(71-1)/2}\\ \\
  &= 0.6470 \; .
\end{align*}%
Remarkably, the Stirling formula approximation for $\PBF_{10}$ differs from the analytic value by only 0.0001.

Finally, we will apply the approximation based on Frame's formula to the \citet{borota2014} summary statistics. This yields%
\begin{align*}
  \PBF_{10} &\approx \Biggr(\frac{8}{2\nu^2-2\nu+1}\Biggr)^{\frac{1}{4}} \cdot \sqrt{\frac{1}{\pi}\Biggl(1+\frac{t^2}{\nu}\Biggr)^{\nu-1}}\\
            &= \Biggr(\frac{8}{2(71)^2-2(71)+1}\Biggr)^{\frac{1}{4}} \cdot \sqrt{\frac{1}{\pi}\Biggl(1+\frac{2.0^2}{71}\Biggr)^{\nu-1}}\\
            &= 0.1684 \cdot 3.8417\\
  &= 0.6469 \; .
\end{align*}%
In this case, the value of $\PBF_{10}$ derived from the Frame approximation is identical to the analytic value to four decimal places.

\section{Comparing the three approximations}

In this section, I compare the accuracies of the three closed form approximation methods (Wendel's approximation, Stirling's formula, and Frame's approximation) for computing the two-sample Pearson Bayes factor. To do this, consider figure \ref{fig:newMethods} below, which plots {\it percent error} as a function of total sample size $N$ (where $N$ ranges from 4 to 100). Here, {\it percent error} is defined as 100 times the absolute value of the difference between the analytic value $C_{\nu}$ and the approximate value of $C_{\nu}$ (which we denote here as $C_{\nu}^*$.), divided by $C_{\nu}$. That is,

\[
  \text{percent error }= 100 \times \frac{|C^*_{\nu}-C_{\nu}|}{C_{\nu}} \; .
  \]

\begin{figure}[t]
  \centering
  \includegraphics[width=0.8\textwidth]{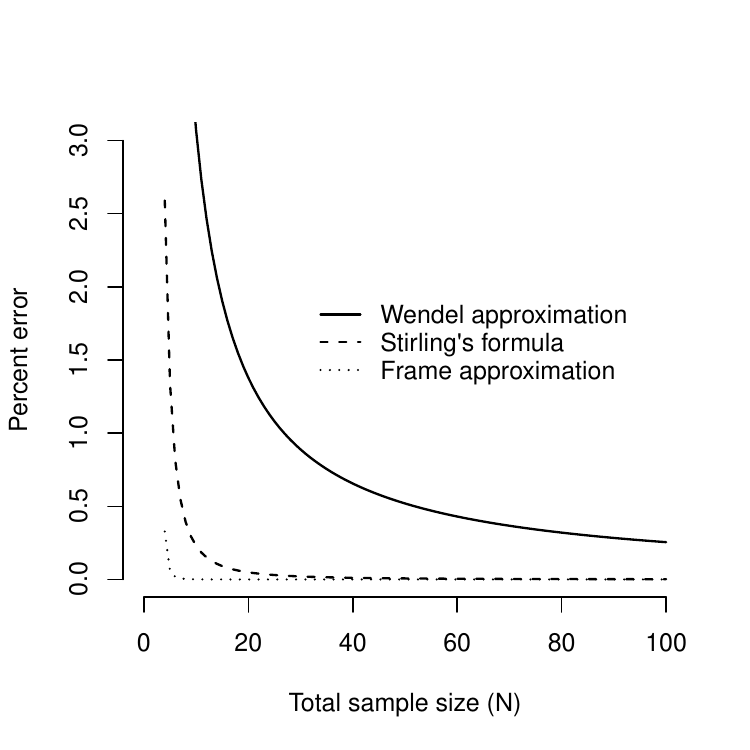}
  \caption{Average percent error of the Wendel, Stirling, and Frame methods (compared to analytic Bayes factor) for values of total sample size $N$ ranging from 4 to 100.}
  \label{fig:newMethods}
\end{figure}

Aligned with our example computations above, Figure \ref{fig:newMethods} shows that all three methods produce quite accurate approximations of the Gamma function quotient $C_{\nu}$ used to compute the Pearson Bayes factor. Moreover, because these methods are asympotic, the approximation gets better as sample sizes increase, which is displayed nicely in the plot. Compared to the Wendel method, the Stirling and Frame methods produce astonishing levels of accuracy, even for small sample sizes. As expected, the Frame quotient method produces the best approximation, with percent error values quickly dropping below 0.01\% for total sample sizes greater than 5. Though less so, the approximation based on Stirling's formula also exhibits similar behavior, with mean percent error values dropping below 0.01\% for total sample sizes greater than 40. Despite the marked differences among the three approaches to approximating the Pearson Bayes factor, the simulation demonstrates what we first observed in our example computations above; all three approaches result in negligible error and are acceptable closed-form approximations to the two-sample Pearson Bayes factor

One may question how these approximations fare compared to the classic BIC method for computing Bayes factors. To answer that question, I conducted a brief simulation study where I compared the approximation error between the analytic Pearson Bayes factor versus (1) the BIC Bayes factor and (2) the approximation error between the {\it worst} performing Gamma quotient approximation (the Wendel method). The choice of comparing against Wendel instead of either Stirling or Frame is based on the idea that any performance gains realized with the Wendel method would be even more improved by using either the Stirling or Frame approximations.

In the simulation, I generated random datasets that each reflected the two-sample designs that we have discussed throughout this paper. For each possible value of $N$ between 4 and 100, I performed 1000 iterations of the following procedure:
\begin{enumerate}
\item Randomly select an ``effect size'' $d$ from a uniform distribution bounded between 0 and 1;
\item The first sample is constructed by randomly drawing $n=\lceil N/2\rceil$ values from a normal distribution with mean 0 and standard deviation 1;
\item The second sample is constructed by randomly drawing $n=\lfloor N/2\rfloor$ values from a normal distribution with mean $d$ and standard deviation 1;
\item Perform an independent samples $t$-test on the means of sample 1 and sample 2, retaining the test statistic ($t$) and the associated degrees of freedom $\nu = N-2$;
\item Using the stored values of $t$ and $\nu$, compute the BIC Bayes factor using the method of \citet{faulkenberry2018} and compute the Pearson Bayes factor using Equation \ref{eq:pbf2}, where $C_{\nu}$ is calculated two different ways: 
  \begin{enumerate}
  \item Analytic formula: $\displaystyle{C_{\nu} = \frac{\Gamma\Bigl(\frac{\nu}{2}\Bigr)}{\Gamma\Bigl(\frac{\nu}{2}+\frac{1}{2}\Bigr)}} \; ;$
  \item Wendel's asymptotic formula: $\displaystyle{C_{\nu} = \sqrt{\frac{2}{\nu}}} \; ;$
    \end{enumerate}
    \item Compute the percent error between the analytic value of the Pearson Bayes factor and the value obtained with each of the two approximate methods (BIC and Wendel).
\end{enumerate}

\begin{figure}[t]
  \centering
  \includegraphics[width=0.8\textwidth]{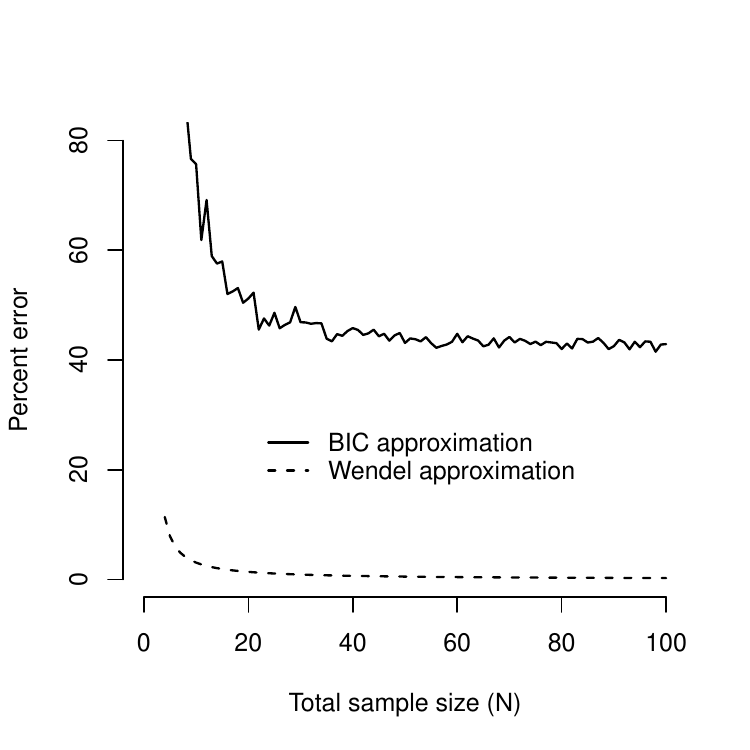}
  \caption{Average percent error of the BIC method and the Wendel method (each compared to the analytic Pearson Bayes factor) for values of total sample size $N$ ranging from 4 to 100.}
  \label{fig:bicCompare}
\end{figure}

The results of the simulation are shown in Figure \ref{fig:bicCompare}. We notice in Figure \ref{fig:bicCompare} that the Wendel formula provides a striking improvement over the BIC method. Whereas the average percent error for the BIC method never gets below 40\%, the average percent error for the Wendel formula approach drops below 1\% as soon as the total sample size reaches 24. As the Stirling and Frame methods provide even better approxmations for $C_{\nu}$ than the Wendel method, it follows that each of the three methods for approximating the two-sample Pearson Bayes factor will provide an immense improvement in calculation accuracy over the classic BIC method.

\section{Conclusion}
In this paper, I have presented three new closed-form approximations of the two-sample Pearson Bayes factor. These techniques allow the user to compute reasonably accurate approximations for Bayes factors in two-sample designs without the need for computing the Gamma function. As such, these computations may be performed using nothing more than a simple scientific calculator, making them a very attractive option for users who wish to compute Bayes factors directly from summary statistics in two-sample designs. Though the formulas vary in complexity, even the simplest formula based on Wendel's (1948) asymptotic formula produces Bayes factor approximations with average percent error dropping below 1\% for reasonably small sample sizes. As all three are asymptotic methods, their relative error will decrease with increasing sample sizes. This is a much better approach to approximating Bayes factors compared to the often-used BIC approximation \citep{kass1995,wagenmakers2007,masson2011,faulkenberry2018,faulkenberry2020,faulkenberry2019ampps}. The approximations presented here retain the spirit of the BIC Bayes factor (e.g., ease of use and ability to compute using only summary statistics), but as demonstrated, they provide a much better level of accuracy. One potential criticism of this new approach is that the presented approximations depend on a particular choice of prior (the Pearson Type VI prior). However, the same is true for the BIC approximation, which also assumes an underlying prior distribution (the unit information prior) \citep{kass1995}. Given that the Bayes factors based on this Pearson Type VI prior show good performance against other well-known Bayes factor techniques \citep{faulkenberry2020}, the approximations presented in this paper are the ideal tool for easily computing evidential value in two-sample designs.

\bibliographystyle{apalike}
\bibliography{references}

\begin{thebibliography}{}

\bibitem[Borota et~al., 2014]{borota2014}
Borota, D., Murray, E., Keceli, G., Chang, A., Watabe, J.~M., Ly, M., Toscano,
  J.~P., and Yassa, M.~A. (2014).
\newblock Post-study caffeine administration enhances memory consolidation in
  humans.
\newblock {\em Nature {N}euroscience}, 17(2):201--203.

\bibitem[Faulkenberry, 2020a]{faulkenberry2020}
Faulkenberry, T. (2020a).
\newblock Estimating bayes factors from minimal summary statistics in repeated
  measures analysis of variance designs.
\newblock {\em Advances in Methodology and Statistics}, 17(1).

\bibitem[Faulkenberry, 2018]{faulkenberry2018}
Faulkenberry, T.~J. (2018).
\newblock Computing {B}ayes factors to measure evidence from experiments: {A}n
  extension of the {BIC} approximation.
\newblock {\em Biometrical Letters}, 55(1):31--43.

\bibitem[Faulkenberry, 2019]{faulkenberry2019ampps}
Faulkenberry, T.~J. (2019).
\newblock Estimating evidential value from analysis of variance summaries: A
  comment on {Ly} (2018).
\newblock {\em Advances in Methods and Practices in Psychological Science},
  2(4):406--409.

\bibitem[Faulkenberry, 2020b]{faulkenberryPearson}
Faulkenberry, T.~J. (2020b).
\newblock The pearson bayes factor: An analytic formula for computing
  evidential value from minimal summary statistics.

\bibitem[Faulkenberry, 2025]{faulkenberryBayes}
Faulkenberry, T.~J. (2025).
\newblock {\em Bayesian Statistics: The Basics}.
\newblock Routledge, New York.

\bibitem[Frame, 1949]{frame}
Frame, J.~S. (1949).
\newblock An approximation to the quotient of {G}amma functions.
\newblock {\em American Mathematical Monthly}, 56:529--535.

\bibitem[García-Donato and Sun, 2007]{gds}
García-Donato, G. and Sun, D. (2007).
\newblock Objective priors for hypothesis testing in one-way random effects
  models.
\newblock {\em Canadian Journal of Statistics}, 35(2):303--320.

\bibitem[Jameson, 2015]{jameson}
Jameson, G. J.~O. (2015).
\newblock A simple proof of {S}tirling's formula for the gamma function.
\newblock {\em The Mathematical Gazette}, 99:68--74.

\bibitem[Jeffreys, 1961]{jeffreys1961}
Jeffreys, H. (1961).
\newblock {\em The {T}heory of {P}robability (3rd ed.)}.
\newblock Oxford University Press, Oxford, UK.

\bibitem[Kass and Raftery, 1995]{kass1995}
Kass, R.~E. and Raftery, A.~E. (1995).
\newblock Bayes factors.
\newblock {\em Journal of the American Statistical Association}, 90(430):773.

\bibitem[Masson, 2011]{masson2011}
Masson, M. E.~J. (2011).
\newblock A tutorial on a practical {B}ayesian alternative to null-hypothesis
  significance testing.
\newblock {\em Behavior {R}esearch {M}ethods}, 43(3):679--690.

\bibitem[Schwarz, 1978]{schwarz1978}
Schwarz, G. (1978).
\newblock Estimating the dimension of a model.
\newblock {\em The Annals of Statistics}, 6(2):461--464.

\bibitem[Wagenmakers, 2007]{wagenmakers2007}
Wagenmakers, E.-J. (2007).
\newblock A practical solution to the pervasive problems of $p$ values.
\newblock {\em Psychonomic {B}ulletin {\&} {R}eview}, 14(5):779--804.

\bibitem[Wang and Liu, 2016]{wang2016}
Wang, M. and Liu, G. (2016).
\newblock A simple two-sample {B}ayesian $t$-test for hypothesis testing.
\newblock {\em The American Statistician}, 70(2):195--201.

\bibitem[Wang and Sun, 2014]{wangSun}
Wang, M. and Sun, X. (2014).
\newblock Bayes factor consistency for one-way random effects model.
\newblock {\em Communications in Statistics - Theory and Methods},
  43(23):5072--5090.

\bibitem[Wasserstein and Lazar, 2016]{asa}
Wasserstein, R.~L. and Lazar, N.~A. (2016).
\newblock The {ASA} statement on $p$-values: {C}ontext, process, and purpose.
\newblock {\em The American Statistician}, 70(2):129--133.

\bibitem[Wendel, 1948]{wendel}
Wendel, J.~G. (1948).
\newblock Note on the {G}amma function.
\newblock {\em American Mathematical Monthly}, 55:563--564.

\end{thebibliography}

\end{document}